\documentclass[sigchi]{acmart}

\usepackage{balance}
\usepackage{booktabs} 
\usepackage{bm}
\usepackage{multirow}

\usepackage{setspace}
\usepackage[ruled]{algorithm2e}

\setcopyright{acmlicensed}

\acmDOI{10.1145/3290605.3300802}

\acmISBN{978-1-4503-5970-2/19/05}

\acmConference[CHI 2019] {CHI Conference on Human Factors in Computing Systems Proceedings}{May 4--9, 2019}{Glasgow, Scotland Uk}
\acmBooktitle{CHI Conference on Human Factors in Computing Systems Proceedings (CHI 2019), May 4--9, 2019, Glasgow, Scotland Uk}
\acmYear{2019}
\copyrightyear{2019}
\acmPrice{15.00}

\begin{document}
\title{REsCUE: A framework for REal-time feedback on behavioral CUEs using multimodal anomaly detection}

\author{Riku Arakawa}
\authornote{These authors contributed equally and are ordered alphabetically}
\affiliation{%
  \institution{University of Tokyo}
  \city{Tokyo}
  \country{Japan}
}
\email{arakawa-riku428@g.ecc.u-tokyo.ac.jp}

\author{Hiromu Yakura}
\authornotemark[1]
\authornote{Also with University of Tsukuba, Japan.}
\affiliation{%
  \institution{Teambox Inc.}
  \city{Tokyo}
  \country{Japan}
}
\email{hiromu@teambox.co.jp}


\renewcommand{\shorttitle}{REsCUE: A framework for REal-time feedback on behavioral CUEs}

\begin{abstract}
Executive coaching has been drawing more and more attention for developing corporate managers.
While conversing with managers, coach practitioners are also required to understand internal states of coachees through objective observations.
In this paper, we present \textit{REsCUE}, an automated system to aid coach practitioners in detecting unconscious behaviors of their clients.
Using an unsupervised anomaly detection algorithm applied to multimodal behavior data such as the subject's posture and gaze, REsCUE notifies behavioral cues for coaches via intuitive and interpretive feedback in real-time.
Our evaluation with actual coaching scenes confirms that REsCUE provides the informative cues to understand internal states of coachees.
Since REsCUE is based on the unsupervised method and does not assume any prior knowledge, further applications beside executive coaching are conceivable using our framework.
\end{abstract}

%
%

\ccsdesc[500]{Human-centered computing~Computer supported cooperative work}
\ccsdesc[300]{Human-centered computing~HCI design and evaluation methods}
\ccsdesc[100]{Information systems~Multimedia and multimodal retrieval}

\keywords{Executive Coaching, Nonverbal behavior analysis, Multimodal interaction, Anomaly detection}

\newcommand{\secref}[1]{the ``\nameref{#1}'' section}
\newcommand{\tabref}[1]{\mbox{Table~\ref{#1}}}
\newcommand{\figref}[1]{\mbox{Figure~\ref{#1}}}
\newcommand{\eqnref}[1]{\mbox{Equation~\ref{#1}}}
\newcommand{\algref}[1]{\mbox{Algorithm~\ref{#1}}}

\renewcommand{\arraystretch}{1.1}

\maketitle

\begin{figure}[htb]
    \begin{center}
        \includegraphics[width=0.48\textwidth]{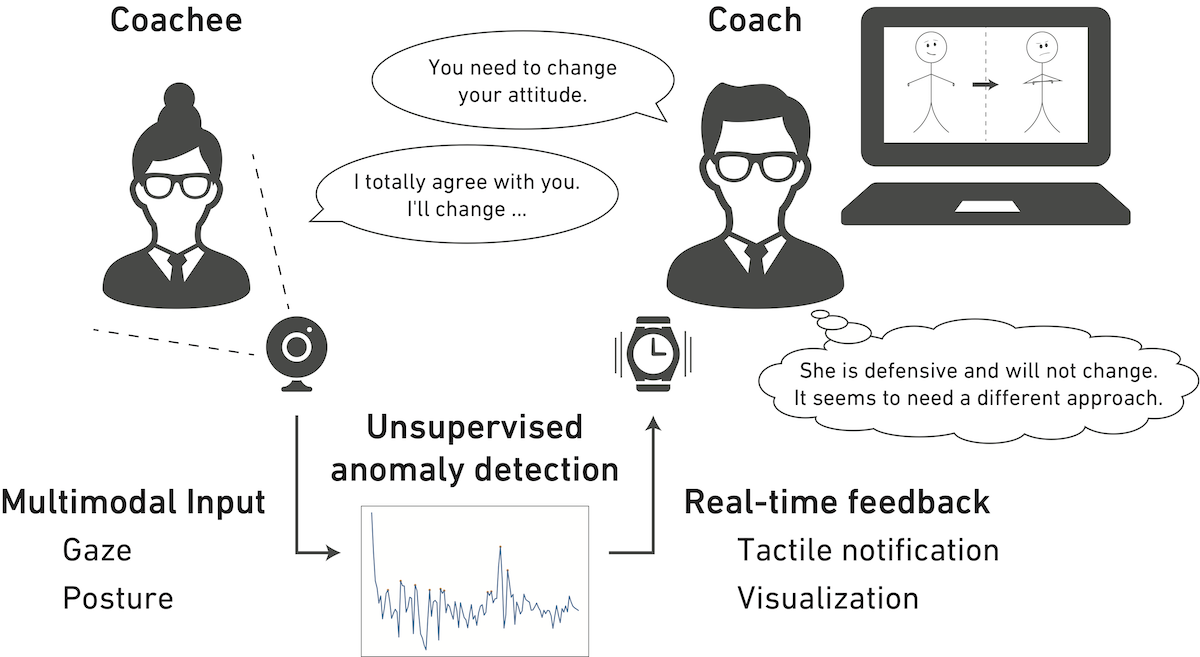}
        \caption{REsCUE detects the behavioral cues of the coachee and notifies the coach in real-time to help the coach understand the internal states of the coachee.}
        \label{fig:usage}
    \end{center}
\end{figure}

\section{Introduction}
\label{sec:introduction}

\textit{Executive coaching} plays an important role in human resource development \cite{doi:10.1037/10355-000,doi:10.1177/0149206305279599}.
As a result, many companies invest in executive coaching to improve the leadership skills or the performances of their managers and the market share of executive coaching has increased to \$2 billion \cite{doi:10.1080/13678860802102534,issn:17502764,doi:10.1016/j.leaqua.2017.11.004}.

Executive coaching usually consists of personal, one-on-one sessions \cite{issn:10559760,doi:10.1037/1065-9293.56.3.154}.
One-on-one sessions are preferred because coaches are required not only to build a rapport with a coachee but also to observe the nonverbal behavior of the coachee during the coaching session \cite{doi:10.1097/01.NAQ.0000264867.73873.1a,isbn:9780761939764}.
For example, the use of disorienting dilemmas is one of an important coaching process \cite{doi:10.1002/ace.20149}; however, sensitive conversations about a coachee's dilemmas may cause deceptive responses \cite{doi:10.1080/10570318809389627}.

In such a situation, coaches are expected to notice a discrepancy between the verbal response and the actual thoughts of the coachee using nonverbal cues \cite{isbn:9781576754924}.
Many articles list the observation skills as one of the skills required for effective coaching, in addition to emotional intelligence and questioning techniques \cite{doi:10.1108/00197850710761945,doi:10.1002/hrdq.1078}.

However, maintaining such objective observations throughout the coaching session requires a great deal of skill \cite{doi:10.1016/j.hrmr.2007.08.008}.
Coaches are often immersed in the verbal communication, paying attention to a deeper or emotional topic or thinking about what to ask next.
In addition, self-deception may interfere with the quality of perception, e.g., ignoring significant behavior unconsciously based on faulty thinking or irrational beliefs \cite{isbn:1446276163}.
Therefore, we expect that the quality of coaching could be improved if the coaches are automatically notified of important nonverbal behavioral cues from the coachee independently of their subjectivity or mental load.

One possible solution is to apply conventional methods proposed in the context of human activity analysis \cite{DBLP:journals/csur/AggarwalR11} or social signal processing \cite{DBLP:journals/taffco/VinciarelliPHPPDS12}.
However, these methods are mainly targeted at classifying human activities into specific categories and therefore have low affinity to the current situation, i.e., each behavior may correspond to various meanings depending on its context in the coaching session \cite{doi:10.1111/1467-6494.00005,doi:10.1037/cpb0000012}.
In addition, most of these methods are designed for post-analysis and are not applicable for providing real-time feedback during a session.
We, therefore, propose \textit{REsCUE}, a new system introducing a real-time anomaly detection method into human behavior analysis (\figref{fig:usage}).

Our framework, by exploiting the anomaly detection method, does not require prior knowledge or heuristic rules and therefore leaves room for the coach's interpretation of the semantics of the behavior based on the context.
Moreover, combined with state-of-the-art behavior analysis methods, our framework is able to detect small but important behavioral cues, which might be missed by the coaches.
REsCUE presents a new perspective of human behavior analysis that augments the perception of the user while leaving its interpretation to the user, which pave the way for further applications in the HCI community.

\subsection{Contribution}

The following four points are the main contributions of this study.

\begin{enumerate}
\item We developed an intelligent system for use in coaching sessions that can automatically detect the nonverbal behavioral cues of coachees and provide feedback to coaches in real-time.
\item Based on a preliminary analysis, we confirmed that the combination of the posture and gaze information is an effective modality for detecting nonverbal cues of coachees.
\item Our user study in actual coaching scenes demonstrates that the proposed system provides informative feedback to professional coaches and would likely improve the quality of sessions.
\item Because the proposed method is based on an unsupervised algorithm and does not assume prior knowledge, it can be applied widely outside of executive coaching applications, as a new framework for real-time behavioral analysis.
\end{enumerate}

\section{Literature Review}

\subsection{Background}

The relation between nonverbal behaviors and internal states has been a distinguished topic in the history of science \cite{book:nonverbal,trove.nla.gov.au/work/11518296}.
Beginning with Charles Darwin \cite{book:darwin}, many researchers have pointed out that nonverbal behaviors are spontaneous and unregulated expressions of internal states \cite{depaulo1992nonverbal}.
On the contrary, the effect of nonverbal behaviors on the internal states has also been revealed, e.g., the influence of facial muscular activity on people's affective responses \cite{Carr5497}.

Based on the relationship, observations of the nonverbal behaviors have been largely focused in various areas including executive coaching, as discussed in \secref{sec:introduction}.
For example, teachers are encouraged to pay attention to nonverbal behaviors of students, which convey their underlying feelings \cite{isbn:0335200001}.
In addition, not only teachers or therapists \cite{isbn:07325223} but also salespeople \cite{isbn:08853134} or entrepreneurs \cite{RePEc:ssi:jouesi:v:4:y:2016:i:2:p:228-239} are expected to get a handle on nonverbal behavioral cues.
Subsequently, a research domain of automatically analyzing nonverbal behaviors, which is often referred to as social signal processing \cite{DBLP:journals/taffco/VinciarelliPHPPDS12}, has spread.

\subsection{Related Work}
\label{sec:related-work}

Many methods have been proposed to analyze human nonverbal behavior using various modalities for one-on-one sessions or group discussions in the context of both human activity analysis \cite{DBLP:journals/csur/AggarwalR11} and social signal processing \cite{DBLP:journals/taffco/VinciarelliPHPPDS12}.
For example, conventional methods relying on handcrafted features have been widely researched, e.g., facial expression recognition \cite{DBLP:conf/fgr/LyonsAKG98,DBLP:journals/ivc/ShanGM09} and posture estimation \cite{DBLP:journals/tsmc/CucchiaraGPV05}.
Conversely, due to the development of deep learning in recent years, end-to-end methods using neural networks have become popular and have shown overwhelming performance improvements.
For example, Wei et al. \cite{DBLP:conf/cvpr/WeiRKS16} achieved state-of-the-art performance in posture estimation by introducing a convolutional neural network.

Based on such analysis technologies, many applications have been proposed \cite{DBLP:journals/taffco/VinciarelliPHPPDS12}.
For example, Sanchez-Cortes et al. \cite{DBLP:journals/tmm/Sanchez-CortesAMG12} proposed a method to detect emergent leaders in a group discussion using handcrafted audio and visual features.
Beyan et al. \cite{DBLP:journals/tmm/BeyanCBM18} applied multiple kernel learning to similar features to predict the leadership styles of emergent leaders.
Hoque et al. \cite{DBLP:conf/huc/HoqueCMMP13} leveraged multimodal behavioral data to generate instructive feedback in the context of training for job interviews.
Nihei et al. \cite{DBLP:conf/icmi/NiheiNT17} introduced a convolutional neural network to extract important utterances from multimodal behavioral data without relying on handcrafted features.
However, these methods are designed to analyze sessions after they occur and are not formulated to provide feedback to coaches in real-time.

Based on the methods to understand human nonverbal behavior in real-time, some studies have proposed systems to provide real-time feedback on social interactions \cite{DBLP:conf/icmi/KuriharaGOMI07,6414963,DBLP:conf/chi/TausczikP13,DBLP:conf/iui/TanveerLH15,DBLP:conf/chi/DamianTBSLA15,DBLP:conf/icmi/SchneiderBRS15,DBLP:conf/mum/MuralidharCNG16}.
For example, Rhema \cite{DBLP:conf/iui/TanveerLH15} is designed to help people with public presentations by providing feedback in real-time via Google Glass based on a speaker's volume and speaking rate.
Logue \cite{DBLP:conf/chi/DamianTBSLA15} addressed a similar situation by providing feedback via head-mounted display based on body energy and openness calculated from hand positions.
For group discussions, Tausczik et al. \cite{DBLP:conf/chi/TausczikP13} proposed a system to analyze the communication patterns of participants and to provide linguistic feedback for improving teamwork.
In addition, Damian et al. \cite{DBLP:conf/icmi/DamianBA16} proposed a general framework to provide real-time feedback on predefined behavioral events represented in an XML format.

These methods are designed to provide explicit feedback based on some specific rules, such as ``louder'' if the speaking voice is faint or ``pay attention to what others are saying'' if the group dynamics are poor.
However, as mentioned in \secref{sec:introduction}, the meaning of the behavior is largely dependent on the context in a coaching session and therefore such explicit feedback would be impossible.

A similar discussion was presented in a proposal of AutoManner \cite{DBLP:conf/iui/TanveerZCTH16}, a system to improve body languages in public speaking, that is ``the appropriateness of the body language is largely dependent on the context of the speech---which is difficult to automatically assess.''
The system overcame this problem by displaying the estimated body skeleton with its changes and distributions in the time series and leaving room for interpretation by the speaker.
However, it is specifically designed for public speaking and therefore not directly applicable to coaching.
Moreover, it assumes the use of post-analysis by speakers themselves and therefore is not able to provide real-time feedback.

\section{Proposed Method}

In this section, we first describe the requirements of the proposed method.
Then, the technical details of the method are presented, including how these requirements are solved.

\subsection{Requirements}
\label{sec:requirements}

To make the coaching sessions more effective using behavior analysis, the following requirements should be considered:

\begin{enumerate}
\item \textbf{Unsupervised detection}~\\
	As discussed in \secref{sec:introduction}, coaches are required to maintain objective and unbiased observations of the coachees.
    That is, assessing the behavior of the coachees based on heuristic rules or human-annotated training data introduces certain criteria and is not appropriate.
    In addition, due to the dependency of the meaning of nonverbal behaviors on their context, designing effective rules or collecting reliable training data is unrealistic.
    Therefore, we need the proposed system to the detect behavioral cues using an unsupervised algorithm.
\item \textbf{Real-time feedback}~\\
	We aim to provide cues for coaches to understand the state of coachees to improve the quality of coaching sessions.
    Therefore, we need the proposed system to detect the cues and provide feedback in real-time, not via post-analysis of a session.
\item \textbf{Intuitive and interpretive feedback}~\\
	We assume that coaches will use the proposed system while conversing with coachees.
    Therefore, the feedback presented to the coaches needs to be intuitive so that they can interpret it at a glance.
    At the same time, feedback that is too abstract, such as presenting only the fact that the behavior has changed or the statistical value of how it has changed, looses its context and is difficult for the coaches to interpret even though it would take a short time to understand.
    Therefore, we need the proposed system to preserve both intuitiveness and interpretiveness with regard to the feedback.
\item \textbf{Non-interruptive notifications}~\\
	Similar to requirement (3), we need to consider how to notify the coaches of the feedback.
    If the notification causes an interruption, e.g., requesting an action by the coach every time, the quality of the coaching session may degrade.
    That is, we need the proposed system to notify in a non-interruptive manner.
\item \textbf{Portable and non-interfering devices}~\\
	Because coaching sessions are often held in the office of the coachee, we need the proposed system to consist of portable devices so that the coach can easily transport them.
    In addition, if the proposed system requires the coachee to wear devices or sensors, this may interfere with their concentration and result in an obstacle to building a rapport.
    Therefore, we also need the system to consist of non-interfering devices.
\end{enumerate}

\subsection{Overview}

\begin{figure*}[tb]
    \begin{center}
        \includegraphics[width=0.85\textwidth]{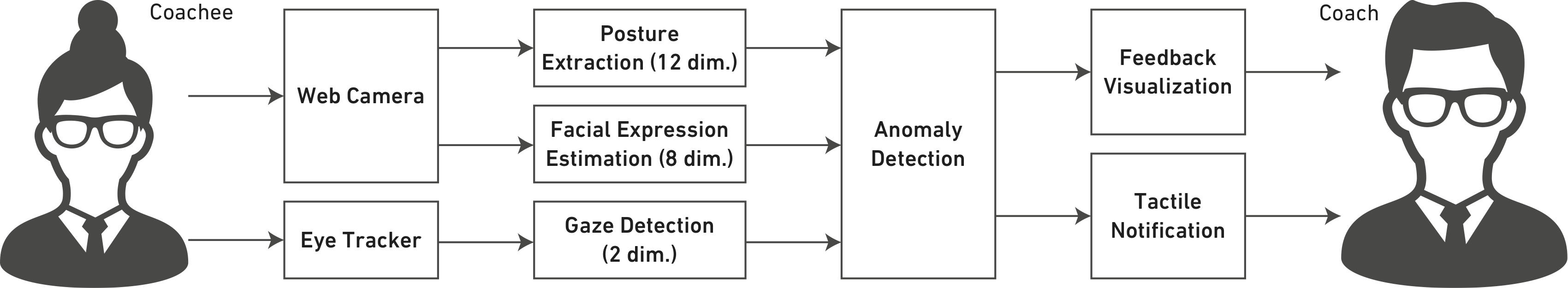}
        \caption{Overview of REsCUE. REsCUE detects important behavioral cues of the coachee via multimodal feature data and notifies the coach both visually and tactilely.}
        \label{fig:overview}
    \end{center}
\end{figure*}

To address the above requirements, we designed the proposed system as shown in \figref{fig:overview}.
The system collects the behavioral data of the coachee via external devices and obtains multimodal feature data.
Using an anomaly detection algorithm, it detects important behavioral cues and notifies the coach both visually and tactilely in real-time.

We now describe the technical specifications of the proposed system along with the rationale behind specification.

\subsection{Multimodal Input}
\label{sec:multimodal}

The proposed system obtains the multimodal behavioral data from the coachees to detect their behavioral cues.

As mentioned in \secref{sec:related-work}, various types of multimodal features have been used in automated behavior analysis.
For example, Nihei et al. \cite{DBLP:conf/icmi/NiheiNT17} leveraged head pose information and speech features and concluded that the combination of these features achieved the best accuracy for detecting important utterances compared to unimodal methods.
Beyan et al. \cite{DBLP:journals/tmm/BeyanCBM18} combined the pose and speech features with the gaze information and reported the effectiveness of the multimodal input.
Hoque et al. \cite{DBLP:conf/huc/HoqueCMMP13} exploited facial expressions for the training of interviewees.

Based on both these studies and the coaching skills mentioned in \secref{sec:introduction}, we prepared three input features: body posture (including head pose), gaze direction, and facial expression.
In detail, along with the fact that the body language captured from the body posture is the basis of the observation skill described in \cite{doi:10.1108/00197850710761945,doi:10.1002/hrdq.1078}, the importance of interpreting the internal state from both the facial expression and the eye of the coachee are emphasized in \cite{isbn:9780761939764}.
Then, we selected which modalities to use in the proposed system later based on a preliminary experiment.

Here, we excluded the speech features due to requirement (4) because it is difficult to construct intuitive and interpretive feedback from changes in speech features. 
While presenting the frequency or the decibel of the coachee's voice is possible, it is difficult to understand how the behavior of the coachee changed and to infer their internal state from such feedback.

In addition, it is possible to use biometric devices to directly capture the signals from the coachee's body.
However, this contradicts requirement (5) and could create a psychological barrier.
Therefore, in this study, we limited the input modalities to those that are measurable without a body-mounted sensor.

\subsubsection{Posture}

The proposed system uses the pose estimation algorithm proposed in \cite{DBLP:conf/cvpr/WeiRKS16}.
Because the algorithm is able to estimate the pose from images taken by a web camera, motion-tracking devices, which contradict requirement (5), are not required.
In addition, the algorithm can not only achieve the state-of-the-art performance, as mentioned in \secref{sec:related-work}, but can also be processed quickly enough to provide real-time feedback, satisfying requirement (2).

\begin{figure}[tb]
    \begin{center}
        \includegraphics[width=0.34\textwidth]{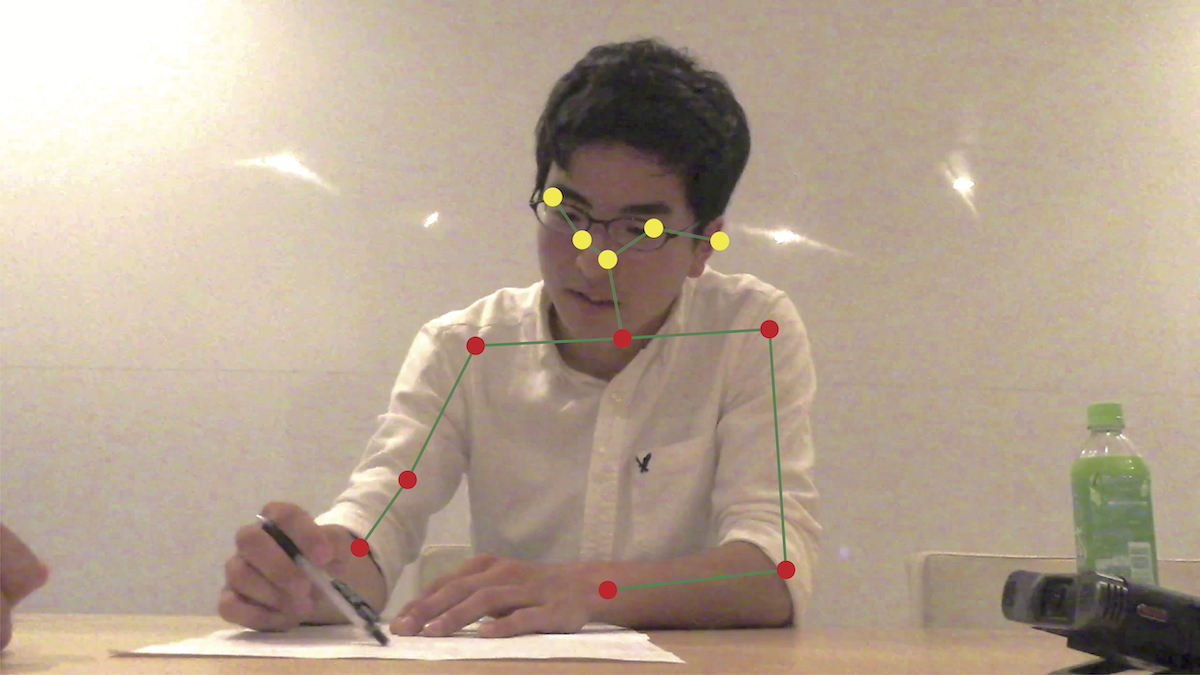}
        \caption{A real example showing the selected key posture points. The yellow circles represent five key points from the head: the nose, both eyes, and both ears. The red circles represent seven key points from the body: the neck, both shoulders, both elbows, and both wrists.}
        \label{fig:openpose}
    \end{center}
\end{figure}

In the proposed system, the 12 key points shown in \figref{fig:openpose} are used to detect the behavioral cues.
We excluded the key points in the lower body because the coaching sessions are usually held at a desk.

\subsubsection{Gaze}

The proposed system uses a commercial eye tracker Tobii 4C to detect the gaze direction.
This is a USB-connected device that is capable of tracking the looking direction without being worn, satisfying requirement (5). 

In the proposed system, we used the two-dimensional coordinate of the gaze position.
Here because the proposed system focuses on the relative change in the detected value, not the absolute value, our system does not require a calibration step for each session.

\subsubsection{Facial Expression}

The proposed system uses MicroExpNet \cite{DBLP:journals/corr/abs-1711-07011} to extract the facial expressions of the coachees.
This is a small and fast convolutional neural network designed for facial expression recognition, which is obtained by distilling a heavy and accurate neural network.
{\c{C}}ugu et al. \cite{DBLP:journals/corr/abs-1711-07011} reported that the network achieved over 95.0\% classification accuracy for the eight expressions of ``neutral,'' ``anger,'' ``contempt,'' ``disgust,'' ``fear,'' ``happy,'' ``sadness,'' and ``surprise'' under the real-time conditions.
Therefore, we decided to use this network to meet requirement (2).

In the proposed system, we use the output value of the final layer after the softmax activation as an eight-dimensional feature vector of the facial expression in the same manner as \cite{7725672}.
In other words, each value of the vector represents the probability that the expression belongs to the corresponding class.

\subsection{Anomaly Detection}
\label{sec:anomaly-detection}

The proposed system detects behavioral cues from the multimodal feature data obtained by the method described in \secref{sec:multimodal}.
To satisfy requirements (1) and (2), we used anomaly detection algorithms, which are sometimes referred to as change point detection algorithms.
This is because there are many proposed unsupervised online algorithms for anomaly detection \cite{DBLP:journals/csur/ChandolaBK09}.

We use SmartSifter \cite{DBLP:journals/datamine/YamanishiTWM04}, an adaptive anomaly detection algorithm based on the Gaussian Mixture Model (GMM).
The main reason we chose the algorithm is that it is one of the most popular unsupervised online anomaly detection algorithms available.
In addition, the results obtained with this GMM-based approach are useful for designing informative feedback, as described later in this section.

Each time new input data arrive, SmartSifter estimates the data's outlierness based on the likelihood calculated by the GMM and, at the same time, updates the parameters of the GMM to fit the input data at the same time.
More formally, let $\bm{x}^{(t)}$ be an input data and $c_i^{(t)}$, $\bm{\mu}_i^{(t)}$, and $\bm{\Sigma}_i^{(t)}$ be the weight, mean, and covariance of the $i$-th component of the $l$-components GMM, respectively, at time $t$.
Then, the outlierness of $\bm{x}^{(t)}$ is calculated as follows.
\begin{equation}
	\label{eqn:outlierness}
	a^{(t)} = - \ln{ \Sigma_{i=0}^{l} \ c_i^{(t - 1)} \mathcal{N} \left( \bm{x}^{(t)} \mid \bm{\mu}_i^{(t - 1)}, \bm{\Sigma}_i^{(t - 1)} \right) }
\end{equation}
The parameters of the GMM are updated subsequently as follows.
\begin{eqnarray}
	\label{eqn:gmm-parameters}
	\gamma_i^{(t)} & = & \frac{c_i^{(t - 1)} \mathcal{N} \left( \bm{x}^{(t)} \mid \bm{\mu}_i^{(t - 1)}, \bm{\Sigma}_i^{(t - 1)} \right)}{\Sigma_{j=0}^{l} c_j^{(t - 1)} \mathcal{N} \left( \bm{x}^{(t)} \mid \bm{\mu}_j^{(t - 1)}, \bm{\Sigma}_j^{(t - 1)} \right)} \nonumber \\
    c_i^{(t)} & = & (1 - r) c_i^{(t - 1)} + r \gamma_i^{(t)} \nonumber \\
    \bar{\bm{\mu}}_i^{(t)} & = & (1 - r) \bar{\bm{\mu}}_i^{(t - 1)} + r \gamma_i^{(t)} \bm{x}^{(t)} \\
    \bm{\mu}_i^{(t)} & = & \bar{\bm{\mu}}_i^{(t)} / c_i^{(t)} \nonumber \\
    \bar{\bm{\Sigma}}_i^{(t)} & = & (1 - r) \bar{\bm{\Sigma}}_i^{(t - 1)} + r \gamma_i^{(t)} \bm{x}^{(t)} {\bm{x}^{(t)}}^T \nonumber \\
    \bm{\Sigma}_i^{(t)} & = & \bar{\bm{\Sigma}}_i^{(t)} / c_i^{(t)} - \bm{\mu}_i^{(t)} {\bm{\mu}_i^{(t)}}^T \nonumber
\end{eqnarray}
Here, $r$ represents a forgetting rate, which is related to the degree of discounting of past input data.

In the proposed system, SmartSifter is extended to use batches in the time series and therefore takes $\bm{X}^{(t)} \in \mathbb{R}^{N \times M}$ as input, where $N$ represents the number of frames in a single batch and $M$ represents the number of the dimensions of the modality data.
This is because the frame-by-frame behavioral changes would include instantaneous physiological responses, which make the feedback to the coach noisy.
Introducing batch processing enables the proposed system to detect the changes in the distribution of the behavior.
This increases the chance of capturing relatively long-term behavioral changes, which may be more difficult to recognize for a human observer \cite{doi:10.1037/a0039166,doi:10.1037/a0014348}.

Moreover, the proposed system exploits SmartSifter to obtain the interpretive feedback, fulfilling requirement (3).
Given that the GMM is used for clustering of the behavioral data, each component of the obtained GMM can be regarded as a representation of a particular behavioral state of the coachee.
Therefore, the proposed system can obtain the representative $l$ frames in a batch at time $t$ as follows.
\begin{eqnarray}
	\label{eqn:representative-frames}
	& \hat{\bm{X}}_i^{(t)} = & \! \! \! \! \bm{X}_{\hat{n}}^{(t)} \ (i = 1, \ldots, l) \\
    & where & \! \! \! \! \hat{n} = \underset{1 \leq n \leq N}{\mathrm{argmax}} \ \mathcal{N} \left( \bm{X}_n^{(t)} \mid \bm{\mu}_i^{(t - 1)}, \bm{\Sigma}_i^{(t - 1)} \right) \nonumber
\end{eqnarray}
Conversely, the most significant outlier frame can be obtained as follows.
\begin{eqnarray}
	\label{eqn:outlier-frame}
	& \check{\bm{X}}^{(t)} = & \! \! \! \! \bm{X}_{\check{n}}^{(t)} \\
    & where & \! \! \! \! \check{n} = \underset{1 \leq n \leq N}{\mathrm{argmin}} \ \Sigma_{i=0}^{l} \ c_i^{(t - 1)} \mathcal{N} \left( \bm{X}_n^{(t)} \mid \bm{\mu}_i^{(t - 1)}, \bm{\Sigma}_i^{(t - 1)} \right) \nonumber 
\end{eqnarray}
By observing both the representative frames from the last batch and the outlier frame from the current batch, the coach can easily understand how the behavior of the coachee has changed while preserving the interpretiveness of the change.

We summarize the above in \algref{alg:anomaly-detection}.
Every time new input data arrive, the outlierness is calculated in the same manner as in \eqnref{eqn:outlierness}.
If the outlierness exceeds the given threshold, the frames obtained using \eqnref{eqn:representative-frames}, which show the representative states so far, and the current outlier frame obtained using \eqnref{eqn:outlier-frame} are presented to the coach as the feedback.
Then, the parameters of the GMM are updated in the same way as in \eqnref{eqn:gmm-parameters}.

\begin{algorithm}[tb]
	\setstretch{1.1}
	\renewcommand{\eqnref}[1]{\mbox{Eq.~\ref{#1}}}
    \caption{The anomaly detection procedure}
    \label{alg:anomaly-detection}
    \SetInd{0.4em}{0.8em}
    \BlankLine
    \KwIn{ \\
        \Indp
        $ l $: the number of components in GMM \\
        $ r $: the forgetting rate \\
        $ a_{th} $: the threshold of the outlierness to give feedback \\
    }
    \BlankLine
    \Begin{
        initialize $ c_i^{(0)}, \bm{\mu}_i^{(0)}, \bar{\bm{\mu}}_i^{(0)}, \bm{\Sigma}_i^{(0)}, \bar{\bm{\Sigma}}_i^{(0)} \ (i = 1, \ldots, l) $ \\
        $ t \gets 1 $ \\
        \While{the new input data $X^{(t)}$ is available}{
          	calculate the outlierness $a^{(t)}$ based on \eqnref{eqn:outlierness} \\
            \If{$ t > 1 $ {\bf and} $ a^{(t)} > a_{th} $}{
            	get the representative frames $\{\hat{\bm{X}}_i^{(t - 1)}\}$ (\eqnref{eqn:representative-frames}) \\
                get the outlier frame $\check{\bm{X}}^{(t)}$ (\eqnref{eqn:outlier-frame}) \\
                give feedback with $\{\hat{\bm{X}}_i^{(t - 1)}\}$ and $\check{\bm{X}}^{(t)}$
            }
            update $ c_i^{(t)}, \bm{\mu}_i^{(t)}, \bar{\bm{\mu}}_i^{(t)}, \bm{\Sigma}_i^{(t)}, \bar{\bm{\Sigma}}_i^{(t)} $ based on \eqnref{eqn:gmm-parameters} \\
            $ t \gets t + 1 $
        }
    }
\end{algorithm}

\subsection{Feedback}
\label{sec:feedback}

To satisfy requirement (4), we make use of tactile feedback to notify the coaches when behavioral cues are detected.
In particular, the proposed system provides coaches with a smartwatch, which vibrates on detection of a behavioral cue.
Tactile feedback is used because it does not interfere with the performance of concurrent tasks while obtaining high notice rates \cite{Sklar2000}.
Moreover, the capability of tactile feedback during social interactions has been confirmed \cite{DBLP:conf/tei/DamianA16}.

At the same time, representative frames from past scenes and the outlier frame from the current scene are displayed to coaches on detection of a behavioral cue.
This allows coaches to easily understand how the behavior has changed, and this information is used to further infer and analyze the internal state of the coachee.
In other words, this visualization achieves both the intuitiveness and interpretiveness as stipulated by requirement (3), because it enables the coaches to grasp feedback at a glance while making the feedback informative.

Moreover, because the scale of body movement patterns varies between individuals \cite{Ekman1980}, it is better to provide the coaches with the capability to adjust the detection threshold during the sessions.
Therefore, we placed ``more'' and ``less'' buttons on the smartwatch to change the threshold $a_{th}$ in \algref{alg:anomaly-detection}. 
In this way, the coaches are able to control the frequency of notifications according to each coachee.

In summary, combining the tactile notification via a smartwatch with intuitive and interpretive visual information, coaches are freed from the burden of paying their attention to displays in parallel to conversing with the coachees.
Moreover, the coaches can easily adjust the sensitivity of the detection on the smartwatch depending on the characteristics of each coachee.
Therefore, the coaches can exploit feedback without losing concentration and the better coaching performances are expected.

\section{Preliminary Experiment}

To determine which modality to use in the proposed system, we conducted a preliminary experiment.
In this section, we describe the procedure and results of the experiment.

\subsection{Data Collection}
\label{sec:data-collection}

The experiment involved three professional coaches (aged 25--39 years old), who participated voluntarily.
Each coach had coaching sessions with two different coachees for at least 30 minutes (4h 28m 35s in total).
All participants, including both the coaches and the coachees, agreed to the use of the collected data for the research purposes.

During the sessions, the behavior of the coachees was recorded with a video camera and a Tobii eye tracker.
After each session, the participating coaches are asked to watch the recorded video and list the top 10 most important behavioral cues of the coachee to infer their internal states.

\subsection{Implementation}

\begin{figure}[tb]
    \begin{center}
        \includegraphics[width=0.40\textwidth]{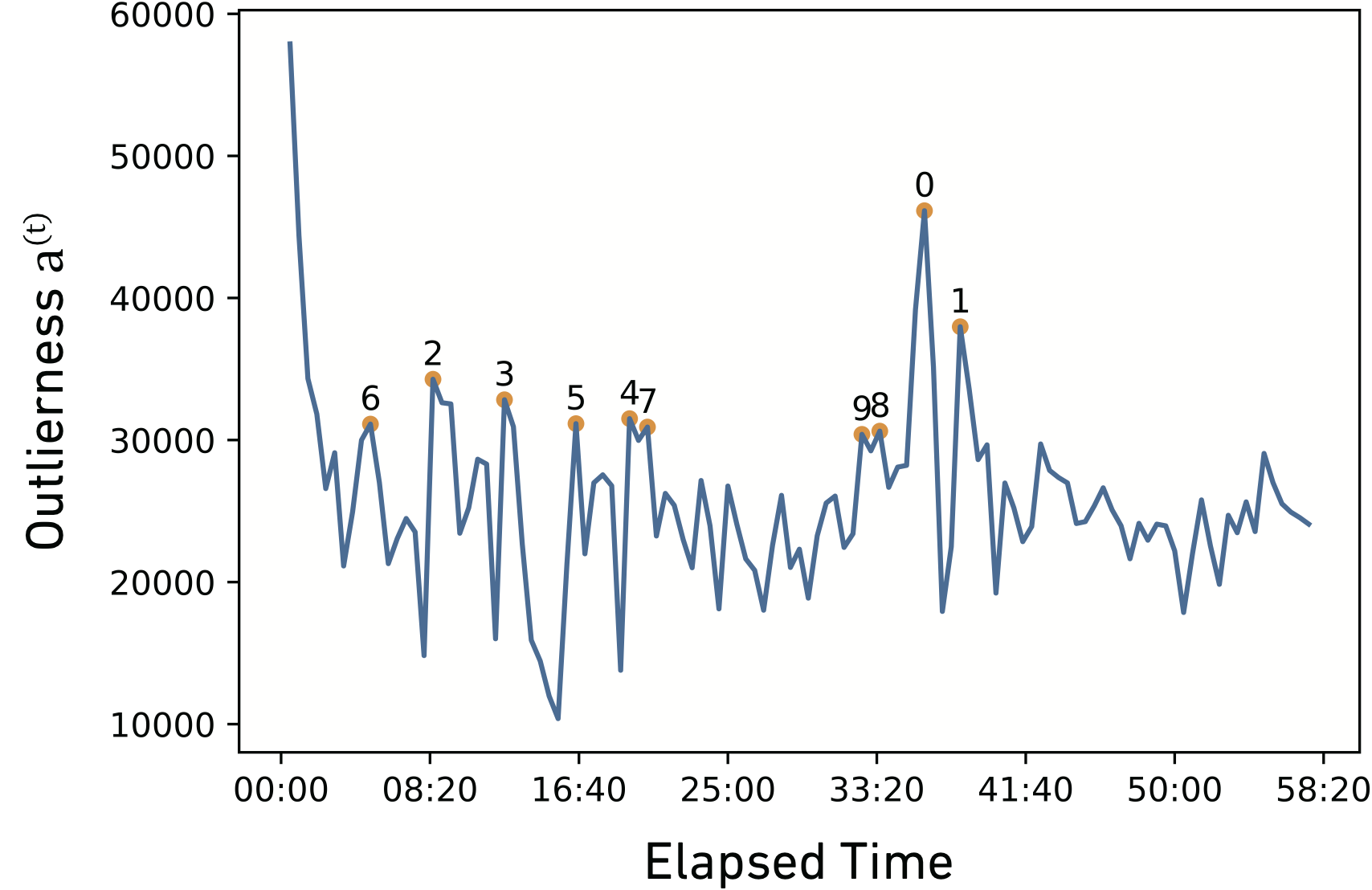}
        \caption{A real example of the transition of the calculated outlierness. The orange points show the detected cues, which are the top 10 peaks of the outlierness.}
        \label{fig:outlier-convergence}
    \end{center}
\end{figure}

We implemented the proposed algorithm and applied it to the recorded data.
Based on our empirical observations, the number of components and the forgetting rate in \algref{alg:anomaly-detection} were set to 2 and 0.1, respectively.
In detail, we found that, as the number of components increased, not only did it take more time until the model's initial convergence, but also it became more difficult for the coach to interpret the detected cues since a larger number of frames were displayed simultaneously.
To compare the detected results with the behavioral cues pointed out by the coaches, we had the system to output the top 10 most significant peaks in the outlierness instead of specifying the threshold.
Here, the first three minutes of each session are excluded for the detection because it takes several minutes for the parameters of the GMM to converge, as shown in \figref{fig:outlier-convergence}.

In addition, we had the proposed system to obtain new behavioral data on every 0.5 seconds and combine them into 30-seconds batches to ensure that the same processing performance was reproducible in a portable computing environment, e.g., a regular laptop with a single GPU, as stipulated in requirement (5).
This is also expected to make the feedback less noisy, as discussed in \secref{sec:anomaly-detection}.

\subsection{Evaluation Metrics}

To compare the effectiveness of each modality, we chose the recall and the minimizing Kendall's $\tau$ distance \cite{DBLP:journals/siamdm/FaginKS03} as evaluation metrics.
The recall represents how many behavioral cues, which are pointed out by the coaches, are captured by the proposed system.
In this case, we set the error tolerance to 30 seconds.
This is because, in addition to the constraint of the batch size, some behavioral cues, such as stretching or scratching one's head, take more than several seconds making it difficult to specify their precise timing.

The minimizing Kendall's $\tau$ distance is a metric to measure the similarity between two top $k$ lists $(k \geq 2)$ and is widely used in the evaluation of search engines \cite{DBLP:journals/tkde/Haveliwala03,DBLP:conf/www/ZieglerMKL05}.
Given the two lists $\tau_1$ and $\tau_2$, the distance is defined as follows.
\[
    K_{min} \left( \tau_1, \tau_2 \right) = \frac{\Sigma_{\left \{ i, j \right \} \in P \left( \tau_1, \tau_2 \right) } \bar{K}_{i, j} \left( \tau_1, \tau_2 \right)}{k \times (k - 1) / 2}
\]
Here, $P \left( \tau_1, \tau_2 \right)$ denotes the set of all unordered pairs of distinct elements in $\tau_1 \cup \tau_2$.
Then, $\bar{K}_{i, j} \left( \tau_1, \tau_2 \right) = 1$ if (i) $i$ appears only in one list and $j$ appears only in the other list, (ii) $i \prec j$ in one list and only $j$ appears in the other list, or (iii) $i \prec j$ in one list and $i \succ j$ in the other list; otherwise, $\bar{K}_{i, j} \left( \tau_1, \tau_2 \right) = 0$.
Consequently, if $\tau_1$ and $\tau_2$ are identical, $K_{min} \left( \tau_1, \tau_2 \right) = 0$.

\subsection{Results}
\label{sec:results}

\begin{table}[tb]
  \caption{The results of the preliminary experiment. The combination of the posture and gaze information showed the highest detection performance.}
  \label{tab:preliminary-result}
  \begin{tabular}{ccc|cc}
    \multicolumn{3}{c|}{Used modalities}        & \multicolumn{2}{c}{Metrics} \\ \hline
    \multirow{2}{*}{Posture} & \multirow{2}{*}{Gaze} & Facial     & \multirow{2}{*}{Recall} & Average of      \\[-0.15em]
                             &                       & expression &                         & $\tau$ distance \\ \hline
    \checkmark &            &            & $0.57 \pm{0.08}$          & $0.42 \pm{0.14}$          \\
               & \checkmark &            & $0.38 \pm{0.15}$          & $0.21 \pm{0.12}$          \\
               &            & \checkmark & $0.08 \pm{0.10}$          & $0.01 \pm{0.02}$          \\
    \checkmark & \checkmark &            & $\mathbf{0.68} \pm{0.08}$ & $\mathbf{0.64} \pm{0.09}$ \\
    \checkmark &            & \checkmark & $0.57 \pm{0.12}$          & $0.40 \pm{0.16}$          \\
               & \checkmark & \checkmark & $0.37 \pm{0.12}$          & $0.20 \pm{0.12}$          \\
    \checkmark & \checkmark & \checkmark & $0.67 \pm{0.16}$          & $0.61 \pm{0.12}$          \\ \hline
  \end{tabular}
\end{table}

The results are shown in \tabref{tab:preliminary-result}.
From the comparison of the recall and the minimizing Kendall's $\tau$ distance, we confirmed that the combination of the posture and gaze information is the most suitable for detecting behavioral cues in coaching sessions.
In addition, as discussed in \secref{sec:multimodal}, the experiment demonstrated the effectiveness of the multimodal features compared to the unimodal features by observing the difference with cases of only the posture or the gaze was used.

The facial expression information, however, did not contribute to improvements in the detection performance.
Here, the chance rate of the recall is $0.11$, meaning that, if we choose the cue points randomly, 1 of the 10 chosen points is considered correct on average.
However, the recall of the case using only the facial expression is lower than that.

\subsection{Analysis}

In this subsection, we discuss the reasons behind the results in \tabref{tab:preliminary-result} by analyzing the detected cues according to each modality.

\subsubsection{Why was the combination of the posture and gaze information effective?}

\begin{figure}[tb]
    \begin{center}
        \includegraphics[width=0.47\textwidth]{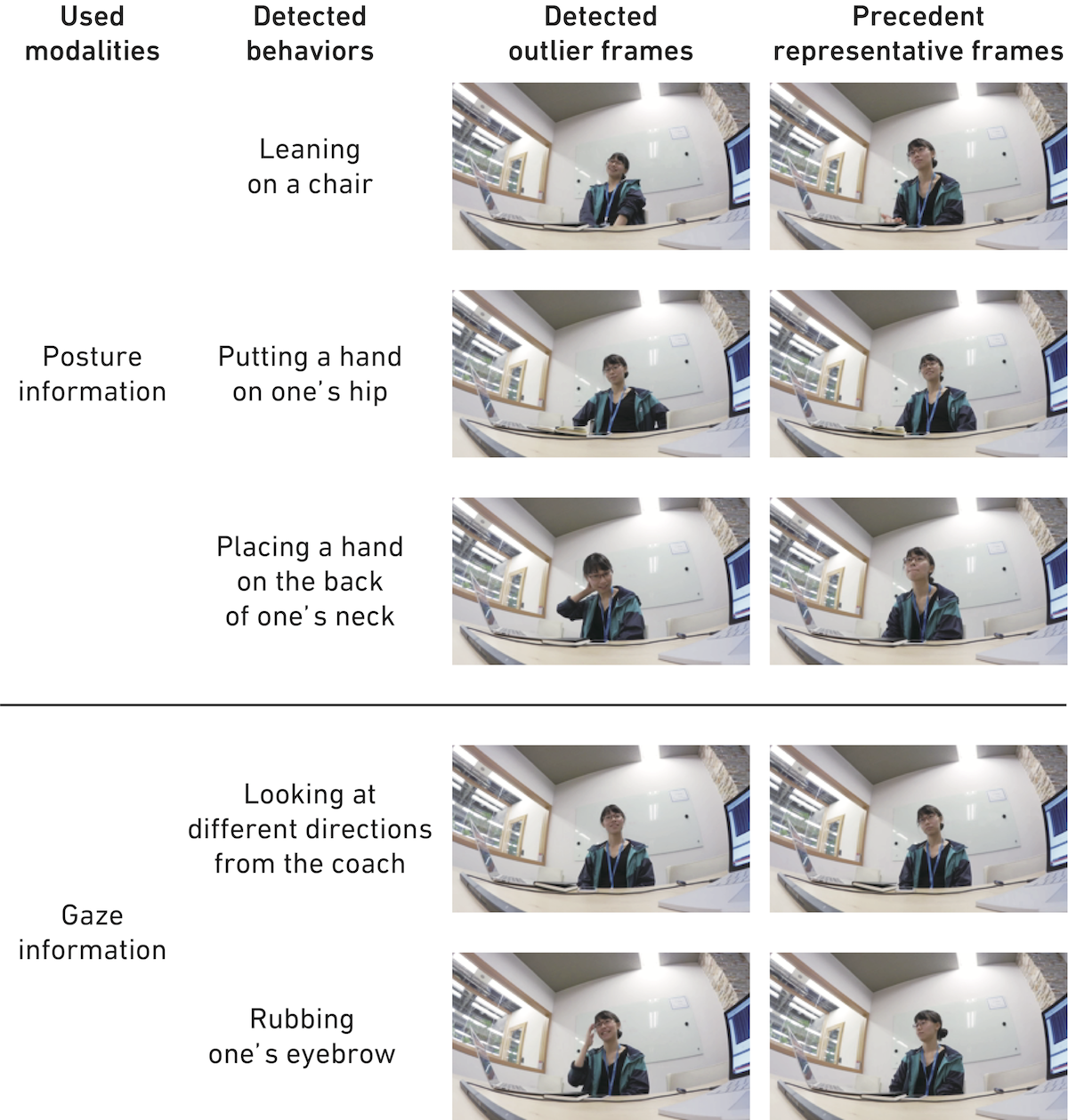}
        \caption{A real example of detected behavioral cues from the one of the recorded sessions. The top three cues are detected using only the posture information, and the next two cues are detected using only with the gaze information. Compared to the preceding representative frames, the changes can easily be seen.}
        \label{fig:preliminary-result}
    \end{center}
\end{figure}

In the recorded data, based on only the posture information, the proposed system detected a wide range of behavioral cues ranging from leaning on a chair to putting a hand on one's hip\footnote{Putting a hand on one's hip is considered to represent a defensive state \cite{doi:10.1155/2012/461247}.} or placing a hand on the back of one's neck\footnote{Placing a hand on the back of one's neck is considered to represent an aggressive state \cite{doi:10.2307/2798193}.}, as shown in \figref{fig:preliminary-result}.
Likewise, the gaze information detected not only changes in the looking direction but also \textit{self-touch} cues \cite{doi:10.2466/pms.101.2.413-420} such as rubbing one's eyebrow\footnote{Rubbing an eyebrow is considered to represent an anxious state \cite{doi:10.2466/pms.101.2.413-420}.}.
This is because self-touch cues often interfere with the detection of the eyes and be captured by the anomaly detection algorithm.
Therefore, the effectiveness of the combination of the posture and gaze information can be attributed to the capability of the system to detect a variety of the behavioral cues.

\subsubsection{Why was the facial expression ineffective?}

As previously stated in \secref{sec:results}, the facial expression did not improve the accuracy of detecting behavioral cues in the sessions.
There appear to be two reasons for this result.
First, facial expressions are obvious and superficial; therefore, the coaches do not regard them as important behavioral cues reflecting the internal states of coachees.
Second, even though the neural network used in the experiment is state-of-the-art, the accuracy might not be sufficiently high.
This is attributable to the fact that MicroExpNet is designed not for faces in free conversation but for posed faces.
Therefore, for example, a face with an open mouth may be classified as fearful even though the coachee is just talking.

\subsubsection{What was the difference between the coaches and the proposed system?}
\label{sec:analysis-difference}

From the results, at least 30\% of the detected points were not included in the behavioral cues listed by the coaches.
When we showed such points to the participating coaches, the points were roughly divided into two groups based on their responses.
The first was a group of obvious and non-informative points such as opening one's notebook or sneezing.
It is because the proposed method uses an unsupervised algorithm and cannot take the meanings of the detected points into consideration.
This result suggests the importance of designing non-interruptive notification, as in requirement (4), so that the coach can easily ignore feedback when it is non-informative.

The other group included points that the coaches agreed were informative.
One participant said: ``Although I did not notice this when I watched the recorded video, once the system pointed out that the coachee had bent slightly forward, I could see that he was opening his mind from about that moment.''
This comment suggests that the proposed system could contribute to improving the quality of coaching sessions by providing the feedback in real-time.

\section{User Study}
\label{sec:user-study}

To confirm the effectiveness of the proposed system, we conducted a user study.
In this section, we describe its setting and results.

\subsection{Implementation}

We implemented a complete version of the proposed system to perform a user study.
The same parameters as \secref{sec:data-collection} were used except for the detection threshold.
The system uses the outlierness of the first peak after three minutes from the beginning of each session as the initial threshold and allows coaches to adjust the threshold subsequently via their smartwatches, as mentioned in \secref{sec:feedback}.

\begin{figure}[tb]
    \begin{center}
        \includegraphics[width=0.36\textwidth]{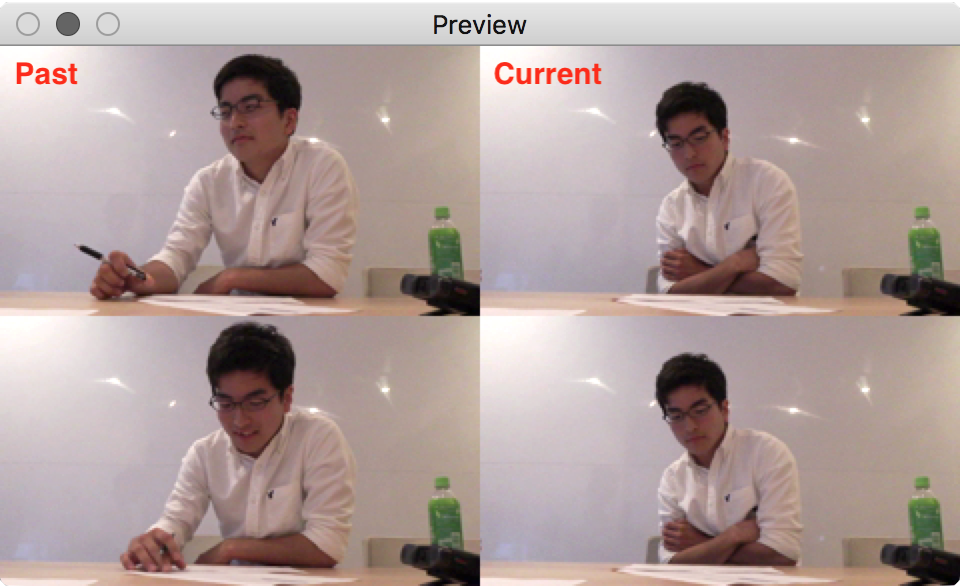}
        \caption{The interface displaying the feedback. The left half presents representative frames from past scenes, and the right half presents the top two outlier frames from current scenes.}
        \label{fig:feedback-display}
    \end{center}
\end{figure}

\begin{figure}[tb]
    \begin{center}
        \includegraphics[width=0.09\textwidth]{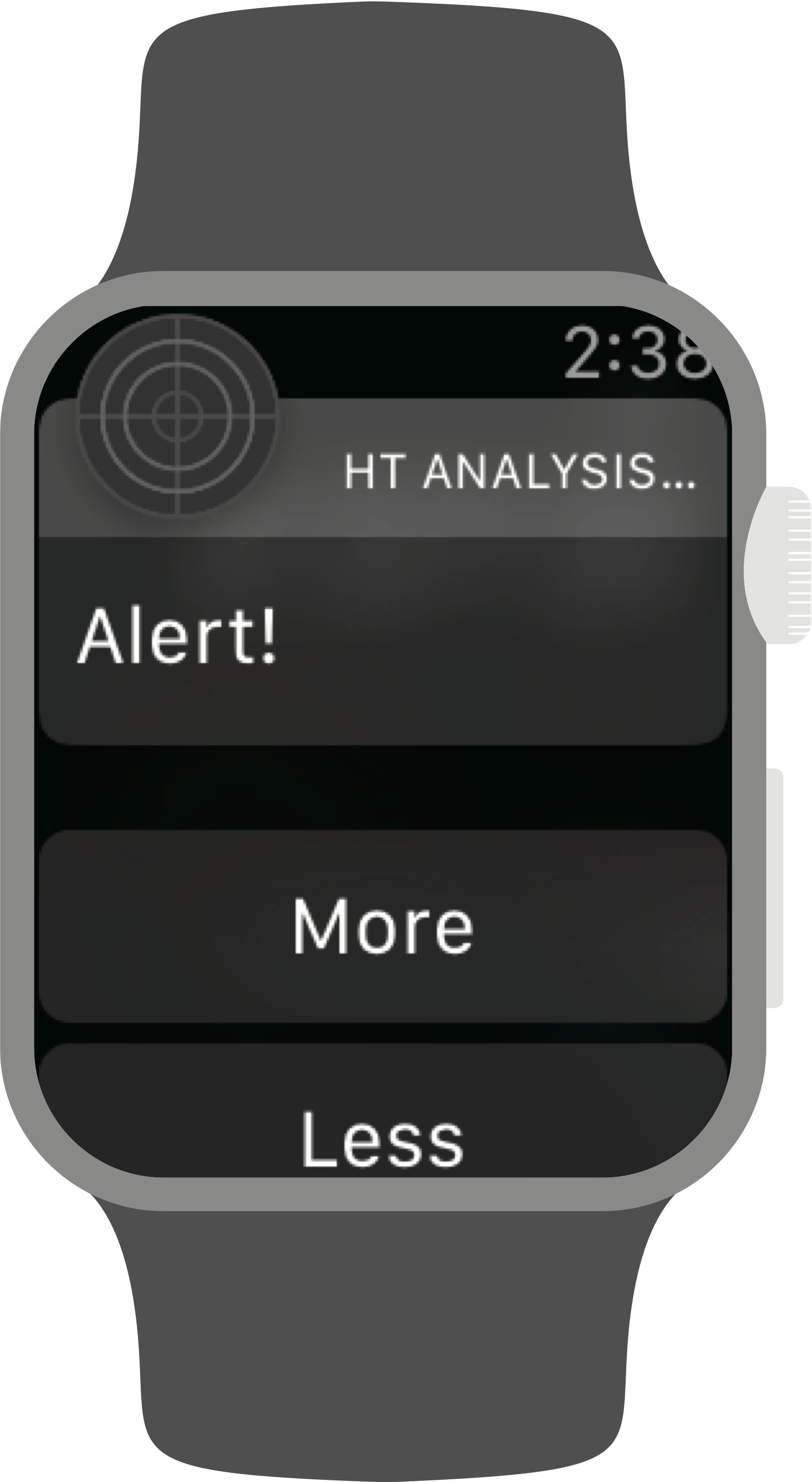}
        \caption{The interface of the control buttons shown on the smartwatch. The coach can adjust the threshold for detecting behavioral cues by pressing these buttons.}
        \label{fig:feedback-watch}
    \end{center}
\end{figure}

The feedback indicating the behavioral cues is presented in \figref{fig:feedback-display}.
It displays the representative frames from past scenes and outlier frames from current scenes so that the coach can understand how the behavior of the coachee has changed.
At the same time, the coach's smartwatch vibrates and shows buttons to adjust the threshold as shown in \figref{fig:feedback-watch}.
Pressing the ``more'' button decreases the threshold while pressing the ``less'' button increases the threshold.

\subsection{Procedure}

\begin{figure}[tb]
    \begin{center}
        \includegraphics[width=0.44\textwidth]{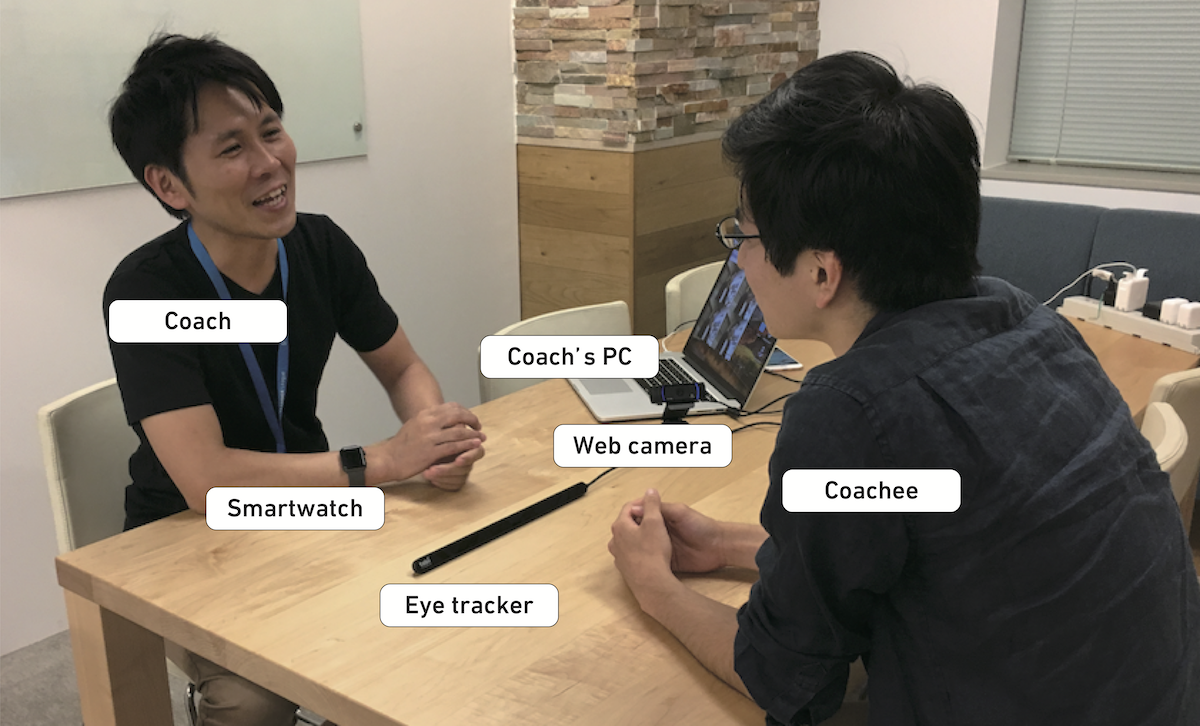}
        \caption{An example of the setup for the user study. The behaviors of the participating coachees were tracked using a web camera and an eye tracker. The participating coaches used a laptop and a smartwatch to receive the feedback.}
        \label{fig:user-study}
    \end{center}
\end{figure}

In the user study, five professional coaches (aged 25--39 years old), including the three coaches from the preliminary experiment, participated voluntarily.
Each coach had coaching sessions with three different coachees (15 sessions in total) using the proposed system, as shown in \figref{fig:user-study}.

Then, we conducted short interviews with the participating coaches.
In this interview, we asked for their subjective opinions concerning the usability of the proposed system.
We also asked whether the given feedback effectively improved the quality of the sessions.
If the participant agreed on the effectiveness, we asked how the sessions changed due to the feedback.

\subsection{Comments}

We found all of the participating coaches responded positively about the proposed system in the subjective interview.
Here, we separately examine the obtained comments concerning the usability and the effectiveness.

\subsubsection{The usability of the proposed system}
\label{sec:comments-usability}

When we asked about the usability of the proposed system, the replies were affirmative, such as ``There was nothing confusing or difficult to understand.'' and ``It was so easy to use that I can imagine that I am using it from tomorrow.''

More specifically, concerning the visualization of the feedback, one participant responded,
\begin{quote}
	``Putting the past frames side by side makes the changes in the behavior obvious.''
\end{quote}
Another participant commented on the comparison with the explicit feedback:
\begin{quote}
	``Simpler feedback such as just showing `defensive' or `opening one's heart' could also be easy to understand. However, if it contradicts my assumptions, I could get confused and might ignore the feedback. In that respect, this system passes the initiative to me and does not cause such confusion while reminding me of other possibilities.''
\end{quote}

Concerning the tactile notification, one participant responded,
\begin{quote}
	``I think the notification is very good because it does not break my concentration and it is not noticed by the subject.''
\end{quote}
From a different perspective, another participant commented on the benefit of the tactile notification:
\begin{quote}
	``When having sessions with about seven or eight clients a day, I sometimes feel out of it. At such a time, the tactile feedback would help me focus on the sessions.''
\end{quote}

The above comments support the usability of the proposed system, as well as the suitability of the feedback design.

At the same time, one participant gave us suggestions for future improvements.
He suggested visualizing the trends of the behavioral cues throughout the session, or throughout multiple sessions of the same coachee:
\begin{quote}
	``Further inferences are possible if this shows that the coachee repeats similar behaviors or that the trend in the behavioral cues changes depending on the topic of conversation.''
\end{quote}
This can be accomplished by applying a clustering algorithm in an unsupervised manner.
For example, the algorithm enables the similarity with past scenes to be represented by visualizing the cue that each cluster belongs to in a time series.

Moreover, this could lead the proposed system to determine the non-informative behavioral cues.
In particular, by adding an ``obvious'' button to the smartwatch in the same manner as shown in \figref{fig:feedback-watch}, clusters of the non-informative cues could be identified in an interactive manner.
In this way, the presence of obvious and non-informative cues, which were discussed in \secref{sec:analysis-difference}, can be reduced.
Therefore, we would like to implement this feature in the near future.

\subsubsection{The effectiveness of the proposed system to improve the quality of the sessions}

The participating coaches also commented positively on the informativeness of the detected behavioral cues and the effectiveness of the proposed system in the sessions, for example:
\begin{quote}
	``Although I often immerse myself in the conversation, thanks to this system, I was able to pay attention to the behavior of the client.''
\end{quote}
and
\begin{quote}
	``This system made me realize that I unconsciously missed or ignored many important behavioral cues.''
\end{quote}

Other comments confirmed that the proposed system helped the coaches change the content of the sessions in accordance with the state of the coachees:
\begin{quote}
	``I had been convinced that the coachee was agreeing to my proposal, but from the given feedback, I noticed that it didn't seem true. So, I was able to make a decision to explain my proposal more carefully until he was satisfied.''
\end{quote}
\begin{quote}
	``I was impressed when the smartwatch vibrated immediately after I asked a delving question having butterflies in my stomach. From the feedback, I became convinced that the underlying cause of the current issue lies there, and succeeded in having a deep discussion in a short period of time.''
\end{quote}

In addition, one participant commented on the educational aspect of the proposed system:
\begin{quote}
	``Up to this time, to cultivate observational skills, we had to review the recordings of the sessions of ourselves or observe the sessions by other coaches. However, using this system, it would be possible to learn what sort of behavioral cues should be focused on during a session and reduce the training time.''
\end{quote}

The above comments suggest that the proposed system may effectively improve the quality of the coaching sessions.

\section{Discussion}

Although the proposed system was generally appreciated in \secref{sec:user-study}, there is still room for further exploration.
In this section, we discuss the limitations and future directions of our research.

\subsection{Limitations}

Throughout the preliminary study, the combination of the posture and gaze information was confirmed to be the most effective and thus chosen as the input modalities in the following user study.
Nonetheless, the number of the participants was relatively small to rule out other possibilities.
Additional investigations with other available modalities are desirable to seek for potential combinations.

Also, though the effectiveness of the proposed system is qualitatively supported by the comments from the participated coaches in the user study, a quantitative comparison of the outcome of the coaching sessions with controlled groups is preferred so as to avoid subject biases.
However, the impact of executive coaching is shaped by a variety of factors such as its purpose, length, organizational context, and individual differences \cite{joo2005executive} and evaluating its outcome via randomized controlled experiments is costly \cite{doi:10.1016/j.leaqua.2017.11.004}. 
One possible remedy is to expand the preliminary experiment to support the results from the other aspects, e.g., collecting self-labelled ground-truth data of the internal states from coachees to validate whether the change of their internal states is captured using the proposed system.

\subsection{Future Directions}

Throughout the user study, the effectiveness of the real-time feedback is confirmed.
In particular, changing the direction of the session on the spot based on the detected cues is not achievable using the conventional methods designed for post-analysis, as we mentioned in \secref{sec:related-work}.

On the other hand, though the design of the proposed feedback system is based on the rationale presented in \secref{sec:requirements}, there are other possibilities like those proposed by previous studies.
We would like to explore a better interaction with the coaches such as comparing different ways of presenting cues, or suppressing non-informative notifications using a clustering method, which is discussed in \secref{sec:comments-usability}. 

Exploring cases of further use also remains a promising endeavor. 
Since the proposed method consists of unsupervised learning and does not require any prior knowledge or rules, it could be used to analyze the behavior of people outside coaching sessions. 
For example, REsCUE might be able to assist people working in dementia care, where it is necessary to analyze the behavior of a patient and consider therapeutic approaches \cite{brooker_2003}. 
Moreover, the connection between conversation and structural neural connectivity in children has been elucidated recently \cite{segaran2018lang}, and thus REsCUE would potentially be utilized in early childhood education as well.

\section{Conclusions}

In this study, we introduced REsCUE, an intelligent system for use in coaching sessions that can automatically detect nonverbal behavioral cues of coachees and provide feedback to coaches in real-time. 
Based on a preliminary experiment, the posture and gaze information proved to be effective modalities to detect behavioral cues.
In actual sessions with professional coaches, a number of favorable comments were obtained, indicating that REsCUE can help coaches to maintain a conversation with coachees while simultaneously inferring their internal states.
For future work, we will investigate other applications of REsCUE by exploiting that the proposed method is based on the unsupervised algorithm and does not depend on prior information.

\appendix


\bibliographystyle{ACM-Reference-Format}
\balance
\bibliography{bibliography}

\end{document}